\documentclass{iopart}

\usepackage{graphicx}

\expandafter\let\csname equation*\endcsname\relax
\expandafter\let\csname endequation*\endcsname\relax

\usepackage{amsmath}
\usepackage{amsfonts}
\usepackage[dvips]{epsfig}
\usepackage{bm}

\usepackage[T1]{fontenc}
\usepackage[latin9]{inputenc}
\usepackage{float}
\usepackage{amsmath}
\usepackage{graphicx}
\usepackage{amssymb}


\begin{document}

\title{Topological Wilson-loop area law manifested using a superposition of loops}
\date{\today}

\author{Erez Zohar}
\address{School of Physics and Astronomy, Raymond and Beverly Sackler
Faculty of Exact Sciences, Tel-Aviv University, Tel-Aviv 69978, Israel.}
\author{Benni Reznik}
\address{School of Physics and Astronomy, Raymond and Beverly Sackler
Faculty of Exact Sciences, Tel-Aviv University, Tel-Aviv 69978, Israel.}

\begin{abstract}
We introduce a new topological effect involving interference of  two meson loops, manifesting a path-independent topological area dependence. The effect also draws a connection between quark confinement, Wilson-loops and topological interference effects.  Although this is only a gedanken experiment in the context of particle physics, such an experiment may be realized and used as a tool to test confinement effects and phase transitions in quantum simulation of dynamic gauge theories.

\end{abstract}

\maketitle

\section{Introduction}

Topological and geometric effects are fundamental quantum-mechanical phenomena. They appear in various physical contexts, as the Aharonov-Bohm and Aharonov-Casher effects \cite{Aharonov1959,Aharonov1984}, Berry's phase \cite{Berry1984} and other models. Such effects are manifested by accumulated topological or geometrical path-dependent phases,
which are observed in interference experiments. Such phases have been experimentally detected several times over the years, and are recently one of the interests of quantum simulations \cite{CiracZoller}; for example, several proposals and experiments probing the effects of an external vector-potential have been suggested with ultracold atoms in optical lattices \cite{Lewenstein2012}.
In quantum field theory and particle physics, a topological phase similar to the Aharonov-Bohm effect appears in the context of the Wilson-loop operator \cite{Wilson,Fradkin1978}, which is an order parameter manifesting the disorder in the confining phase of a gauge theory. The Wilson-loop operator along a curved spacetime path $C$ is \footnote{For the sake of simplicity, we use abelian terms. The non-abelian generalization is straightforward and can be found, for example, in \cite{PolyakovBook}}
\begin{equation}
W \left( C \right) = P\left(e^{i \oint_{C}A_{\mu}dx^{\mu}} \right)
\end{equation}
where $P$ stands for path ordering (see , for example, \cite{PolyakovBook}).

Wilson-loops manifest confinement through the area dependence of their expectation value, and thus are an important test for confinement. They are extremely useful in Euclidean spacetime, for numerical (Monte-Carlo) simulations.
 In Minkowski spacetime, besides the fact they involve a product of operators along a loop and thus are non-local, they can be interpreted as transition amplitudes, and thus their phases do not contribute to the related probabilities.
 To gain information from these phases, one has to use interference effects as we propose here.

In this paper, we present a method to observe the area-law manifested in confining theories using superposition and interference of mesons, which unlike in the Wilson-loop approach, contains the relevant phase as a relative one. We draw the connection between our method and the well-known Wilson-loop.
Note that previous works have already discussed the properties of the inter-quark potential using multiple Wilson-loops \cite{Bachas1986,Nussinov2001}.
Measurements of Wilson-loop operators were discussed in \cite{Beckman2002}.

The paper is organized as follows: first, in section II, we consider the interference effect of two mesons, consisting of static quarks, in a superposition, and show how to gain the string tension from it. The relation to the Wilson-loop operator is drawn. In section III, we allow one of the quarks to be dynamic, modeling it as a particle in a moving harmonic potential, and show how to obtain the string tension in that case, using an exact solution of Schrodinger equation. We discuss the relation of our approach to ordinary Wilson-loops in section IV.
Finally, in section V, we discuss the possibility of realizing the proposed idea using a quantum simulator \cite{CiracZoller}.

\section{Superposition of loops: static quarks}

Free quarks can not be found in nature, but rather form hadrons, due to the phenomenon of quark confinement \cite{Wilson}. A quark and an anti-quark, attached to each other by a confining flux-tube, form a meson, which is the simplest hadron in QCD. The static potential between the quarks, as a function of their distance $R$, takes the form
\footnote{Generally speaking, $V\left(R\right)$ should also include a $\propto R^{-1}$ term. However, we neglect it, assuming that $R$ is large enough. This is sufficient for the interference experiment proposed hereby as it is only affected by an area difference.}
\begin{equation}
V\left(R\right) = \gamma R
\end{equation}
for large values of $R$, where $\gamma=\gamma(g^2)$ is called the string tension, and $g$ is the coupling constant \cite{Wilson,KogutSusskind,Polyakov,KogutLattice}.

\begin{figure}[!t]
\begin{center}
\includegraphics[scale=0.55]{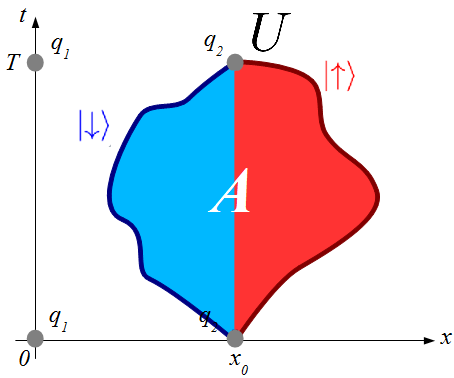}
\end{center}
\caption{The gedanken-experiment. Two fermions $q_1$,$q_2$ form a meson of length $x_0$ at time $t=0$. While $q_1$ remains static, $q_2$ is externally moved (with no dynamics of its own) in a superposition of two possible trajectories, corresponding to its internal states $\left|\uparrow\right\rangle$ and $\left|\downarrow\right\rangle$, creating a superposition of mesons. The path of the $\uparrow$ fermion is drawn in red, and the path of the $\downarrow$ one - in blue. At time $t=T$, both the trajectories arrive again at $x_0$, and the state is then mixed using a unitary transformation (equation (\ref{unit})), causing an interference effect.
In case of confinement, the interference phase depends on the area $A$ (defined in equation (\ref{Adefinition})) in case of confinement - as does the Wilson-loop. The phase is accessible from the probabilities to find $q_2$ in either of its internal states (equation (\ref{staticprobs})). }
\label{fig1}
\end{figure}

Consider two \emph{static} quarks, initially separated by distance $x_0$. By static we mean that we treat them as external sources of electric field, which do not have their own dynamics, but can be externally moved. Taking confinement into account, we know that the two quarks are connected by a long flux-tube, with length $R$, forming a "meson". If we allow one quark to move along some trajectory $x\left(t\right) = x_0 + d\left(t\right)$, we get the Hamiltonian
\begin{equation}
H_{stat} = \gamma \left(x_0 + d\left(t\right)\right)
\end{equation}
where we generally assume that $\gamma$ is an unknown quantity.

 Next, assume that the moving quark has some two internal energy levels $s$, which we denote by $\left|\uparrow\right\rangle,\left|\downarrow\right\rangle$ - eigenstates of the Pauli Matrix $\sigma_z$.
 We define the projection operators to the subspaces of internal levels as $P_\uparrow = \left|\uparrow\right\rangle\left\langle\uparrow\right|,P_\downarrow = \left|\downarrow\right\rangle\left\langle\downarrow\right|$, and introduce an internal level dependent paths $d_s\left(t\right)$, i.e.,
\begin{equation}
H_{stat} = \gamma \left(x_0 + \underset{s}{\sum}d_s\left(t\right)P_s\right)
\end{equation}
The energy levels are used to generate level-dependent spatial positions. The choice of these levels depends on the separation method. For example, if one considers an abelian theory, where $s$ corresponds to the spin ($S_z$), the separation could be achieved using an external magnetic field, $\mathbf{B} = \mathbf{B}\left(x,t\right)\mathbf{\hat z}$, then $\gamma \underset{s}{\sum}d_s\left(t\right)P_s = - \mathbb{\mu} \cdot \mathbf{B}$.  A concrete example of generating such a separation is found in \cite{Dynamic}.

We wish to consider the interference effect of two mesons in superposition, varying their length. We shall consider a superposition of a meson with a $\uparrow$ fermion and a meson with a $\downarrow$ fermion, initially with the same length (i.e., $d_s\left(t\right) \equiv 0$ for $t \leq 0$), which are stretched by moving the right fermion in two opposite directions, and then, at time $T>0$, brought back to the same length (i.e., $d_s\left(t\right) \equiv 0$ also for $t \geq T$. The opposite directions impose more conditions on the paths: $d_\uparrow\left(t\right) \geq 0$, $d_\downarrow\left(t\right) \leq 0$, and Since the $\downarrow$ fermion should not "go through" the static fermion at $x=0$, we also demand $\left|d_\downarrow\left(t\right)\right| < x_0$. The initial state is, of course,
\begin{equation}
\left|\psi\left(0\right)\right\rangle = \left|\uparrow_x\right\rangle = \frac{1}{\sqrt{2}}\left(\left|\uparrow\right\rangle+\left|\downarrow\right\rangle\right)
\end{equation}

Solving Schrodinger equation, one gets that
\begin{equation}
\left|\psi\left(T\right)\right\rangle = \frac{e^{-i \gamma x_0 T}}{\sqrt{2}} \left(e^{-i \gamma \int_0^T d_{\uparrow}\left(t'\right)dt'}\left|\uparrow\right\rangle+e^{-i \gamma \int_0^T d_{\downarrow}\left(t'\right)dt'}\left|\downarrow\right\rangle\right)
\label{statschro}
\end{equation}
and so in order to measure $\gamma$, one has merely to cause an interference between the two states.
We perform Ramsey interference, by applying the rotation $U=e^{-i \frac{\pi}{4} \sigma_y}$ on the state at $t=T$, and one obtains, up to a global phase,
\begin{equation}
U\left|\psi\left(T\right)\right\rangle = \left( \sin \left(\frac{\gamma A}{2}\right)\left|\uparrow\right\rangle + i\cos \left(\frac{\gamma A}{2}\right)\left|\downarrow\right\rangle\right)
\label{unit}
\end{equation}
where
\begin{equation}
A = \int_0^T \left(d_{\uparrow}\left(t'\right) - d_{\downarrow}\left(t'\right)\right)dt'
\label{Adefinition}
\end{equation}
is the area enclosed between the two trajectories (as in figure \ref{fig1}). Thus, the probabilities to find the system on each of the internal levels are
\begin{equation}
P_{\uparrow}=\sin^2\left(\frac{\gamma A}{2}\right) ;
P_{\downarrow}=\cos^2\left(\frac{\gamma A}{2}\right)
\label{staticprobs}
\end{equation}

By performing such an interference experiment, and measuring the phase, knowing the area difference one can calculate the string tension $\gamma$. Moreover, if the phase does not exhibit such an area law, it means that the system is not within a confining phase, and hence this measurement can be used for probing the confining phase as well.

\section{Dynamic quark model}
Next we wish to introduce a simple model of dynamical charges, in order to examine the corrections to the latter static case. In particular, we wish to examine decoherence and destructive interference due to excitations of the mesons. For the sake of simplicity, we assume that one of the quarks is static (the one placed in $x=0$), and that the other one, having a mass $m$, is trapped in a harmonic potential, centered around $x=x_0>0$. This harmonic trap is merely an external trapping potential.
The Hamiltonian of the system takes the form
\begin{equation}
H_0 = \frac{p^2}{2m} + \frac{1}{2}m\omega^2\left(x-x_0\right)^2 + \gamma x
\end{equation}

  Assume that the dynamic quark has two internal energy levels, denoted and treated as before. We introduce an interaction between the internal and external degrees of freedom, of the form
\begin{equation}
H_{int} = x \underset{s}{\sum} G_s\left(t\right)P_s
\end{equation}
% + \underset{s}{\sum} \left(G_s\left(t\right)x_0 - \frac{1}{2 m \omega^2}G^2_s\left(t\right)\right)P_s
Where $G_{s}\left(t\right)$ are opening functions which are zero for times $t \leq 0, t \geq T$, and are assumed to be smooth enough, i.e. at least their first and second time derivatives vanish for $t=0,T$.

Define $d_s\left(t\right) = -\frac{G_s\left(t\right)}{m \omega^2}$ (the same conditions on $d_s\left(t\right)$ apply as in the static case, of course, and that poses conditions on
$G_s\left(t\right)$), $x_s\left(t\right) = x_0 - \frac{\gamma}{m \omega^2} + d_s\left(t\right)$. Then one gets, after completing the square, that the total Hamiltonian is (neglecting constants)
\begin{equation}
H = \underset{s}{\sum}\left(H_s + \gamma d_s\left(t\right) + f_s\left(t\right) \right)P_s
\end{equation}
where the first part is just a Harmonic oscillator, with the center of its potential following classically the trajectory $x_s\left(t\right)$:
\begin{equation}
H_s = \frac{p^2}{2m} + \frac{1}{2}m\omega^2\left(x-x_s\left(t\right)\right)^2
\end{equation}
and $f_s\left(t\right) = G_s\left(t\right)x_0 - \frac{1}{2 m \omega^2}G^2_s\left(t\right)$.

 \subsection{Exact Solution of Schrodinger Equation}
The next step is the solution of Schrodinger Equation for our Hamiltonian:
 \begin{equation}
 i \frac{\partial}{\partial t} \left|\psi\right\rangle = H \left|\psi\right\rangle
 \end{equation}
 Noting that the internal levels are not changed by the Hamiltonian, we can first solve for a given internal level $s$ and consider only the harmonic part. Thus if we set for a given $s$,
 \begin{equation}
 \left|\psi^{\left(s\right)}\right\rangle = e^{-i \left(\gamma \int_0^t d_s\left(t'\right)dt' + \int_0^t f_s\left(t'\right)dt' \right)} \left| \tilde \psi^{\left(s\right)}\right\rangle
  \end{equation}
 we get that $\left| \tilde \psi^{\left(s\right)}\right\rangle$ is the solution of a Schrodinger equation for a classically moving Harmonic potential:
  \begin{equation}
 i \frac{\partial}{\partial t} \left| \tilde \psi^{\left(s\right)}\right\rangle = H_s \left| \tilde \psi^{\left(s\right)}\right\rangle
 \end{equation}
 using the solution of this equation in $x$-space, we get that the solutions (not eigenstates) are
 \begin{equation}
 \psi^{\left(s\right)}_n \left(x,t\right)  = e^{i \tilde \Phi_0^{\left(s\right)} \left(t\right)}
 e^{-i E_n t}
 e^{-i \gamma \int_0^t d_s\left(t'\right)dt'}
 e^{i m \left(\dot d_s \left(t\right)+ \dot q_s \left(t\right)\right) \left(x-x_s\left(t\right)\right)}
 \chi_n\left(x-x_s\left(t\right)-q_s\left(t\right)\right)
 \end{equation}
  where $E_n, \chi_n\left(x\right)$ are the energies and eigenstates of a "regular", fixed-potential harmonic oscillator, $\tilde \Phi_0^{\left(s\right)} \left(t\right) = - \int_0^t f_s\left(t'\right)dt' + \frac{m}{2}\int_0^t \dot d_s^2\left(t\right) - \frac{m}{2}\int_0^t\left( \dot q_s^2\left(t\right) - \omega^2 q_s^2\left(t\right)\right)$ and
  $\ddot q_s + \omega^2 q_s = - \ddot d_s$.

  Our initial condition is $\psi^{\left(s\right)} \left(x,0\right)  = \chi_0 \left(x-x_s\left(0\right)\right) = \chi_0 \left(x-x_0\right)$. For $t \leq 0$, the oscillator is supposed to be in its non moving ground state, and thus we expect that $q_s\left(t\right) \equiv 0$ for these times. Thus $q_s \left(0\right) = 0$. Using the continuity of the equation of motion of $q_s$, we get that $\dot q_s \left(0\right)=0$ as well. From the smoothness of the opening functions we know that $\dot d_s \left(0\right)=0$. Thus we conclude that in order to start from a local ground state, the solution must be
  \begin{equation}
 \psi^{\left(s\right)} \left(x,t\right)  = e^{i \tilde \Phi_0^{\left(s\right)} \left(t\right)}
 e^{-i E_0 t}
 e^{-i \gamma \int_0^t d_s\left(t'\right)dt'}
 e^{i m \left(\dot d_s \left(t\right)+ \dot q_s \left(t\right)\right) \left(x-x_s\left(t\right)\right)}
 \chi_0\left(x-x_s\left(t\right)-q_s\left(t\right)\right)
 \end{equation}

Next, we wish to interpret this solution in terms of local instantaneous eigenvalues. That is, the states $\left|n\left(t\right)\right\rangle$, defined as eigenvalues of $H_s\left(t\right)$ in the "frozen" time $t$. We already know that the system starts at $t=0$ with the eigenstate $\left|0\left(0\right)\right\rangle$, but what's later? In order to do that, we define $y = x - x_s$. Writing $H_s\left(t\right)$in terms of $y$ at a fixed $t$ yields this diagonalization: it is merely a transformation to a frame which moves with the potential, which is its instantaneous rest frame (IRF).
Consider the position-dependent part of $\psi^{\left(s\right)} \left(x,t\right)$, in terms of the IRF. Call it $\phi^{\left(s\right)} \left(y,t\right)$:
\begin{equation}
\left|\phi^{\left(s\right)} \left(t\right)\right\rangle =
 e^{i m \left(\dot d_s \left(t\right)+ \dot q_s \left(t\right)\right) y} e^{-i q_s \left(t\right) p_y} \left|0\left(t\right)\right\rangle
\end{equation}
in the IRF basis, this is a coherent state - Poissonian distribution of $\left|n\left(t\right)\right\rangle$ states:
\begin{equation}
\left|\phi^{\left(s\right)} \left(t\right)\right\rangle = e^{\frac{i}{2}m\left(\dot d_s \left(t\right)+ \dot q_s \left(t\right)\right) q_s\left(t\right)}\left|\alpha_s\left(t\right)\right\rangle
\end{equation}
where
\begin{equation}
\alpha_s \left(t\right) = \sqrt{\frac{m \omega}{2}}q_s\left(t\right) + \frac{i m \left(\dot d_s \left(t\right)+ \dot q_s \left(t\right)\right)}{\sqrt{2 m \omega}}
\end{equation}

 \subsection{Superposition and Coherence}
 Suppose we start, at $t=0$, with an initial state
 \begin{equation}
 \left|\psi\left(0\right)\right\rangle =  \left|0\left(0\right)\right\rangle \left|\uparrow_x\right\rangle =
 \frac{1}{\sqrt{2}}  \left|0\left(0\right)\right\rangle \left( \left|\uparrow\right\rangle +  \left|\downarrow\right\rangle \right)
 \end{equation}
 then, defining $\Phi^{\left(s\right)}_0\left(t\right) = \tilde \Phi^{\left(s\right)}_0\left(t\right) + \frac{1}{2}m\left(\dot d_s \left(t\right)+ \dot q_s \left(t\right)\right) q_s\left(t\right)$ and using the solutions from the previous section, we get
 \begin{equation}
 \left|\psi\left(t\right)\right\rangle =
 \frac{1}{\sqrt{2}} e^{-i \omega t /2} \underset{s}{\sum}e^{i \Phi^{\left(s\right)}_0\left(t\right)} e^{-i \gamma \int_0^t d_s\left(t'\right)dt'} \left|\alpha_s\left(t\right)\right\rangle \left|s\right\rangle
 \end{equation}

 Let us understand the meaning of this state. We start at $t=0$ with a superposition of two states with two different values of $s$. Both of them are in the ground state of an oscillator, centered around the same position ($x_+\left(0\right) = x_-\left(0\right) = x_0$). Then we "move" the wavefunctions together with the potential: the opening functions are translated to the trajectories $d_s \left(t\right)$. Each element of the superposition "goes" through another path, since the moving of the potential depends on the internal level. A superposition of coherent states in the terms of the local IRFs is created, and eventually, at $t=T$, both the interaction functions are closed, i.e. the two potentials experienced by internal levels overlap again, and $x_+\left(T\right) = x_-\left(T\right) = x_0$.

 Next, trace out the oscillator degrees of freedom at $t=T$, to obtain an internal-level density matrix. The density matrix at $t=T$ is
 \begin{equation}
\rho =
 Tr_{\text{osc}} \tilde \rho
= \frac{1}{2} \begin{pmatrix}
1 & e^{i\left(\Phi^{\left(\uparrow\right)} - \Phi^{\left(\downarrow\right)}\right)}B \\
e^{-i\left(\Phi^{\left(\uparrow\right)} - \Phi^{\left(\downarrow\right)}\right)}B^* & 1
\end{pmatrix}
 \end{equation}
 where $\tilde \rho = \left|\psi\left(T\right)\right\rangle  \left\langle\psi\left(T\right)\right|$, $\Phi^{\left(s\right)} = \Phi^{\left(s\right)}_0\left(T\right)-\gamma \int_0^T d_s\left(t'\right)dt'$
 and $B = e^{-\frac{1}{2}\left(\left|\alpha_\uparrow\left(T\right)\right|^2 + \left|\alpha_\downarrow\left(T\right)\right|^2\right)}e^{\alpha_\uparrow\left(T\right)\alpha_\downarrow^*\left(T\right)}$.

One can see that the phases depend on two parts: one, $\Phi^{\left(s\right)}_0\left(T\right)$, is totally calculable. The other is $\gamma$ dependent and can't be calculated unless $\gamma$ is known. The $\gamma$ dependent phases are only global (the probabilities - the diagonal terms in $\rho$ do not depend on them). In order to observe the phases, we wish to cause an interference, i.e. to rotate the state: We act on the system at $t=T$ with the rotation operator $U=e^{-i \frac{\pi}{4} \sigma_y}$.
One can use the diagonal terms in the new density matrix $U \rho U^{\dagger}$ to determine the probabilities to measure each of the internal levels. The probabilities are \begin{multline}
P_\uparrow = \frac{1}{2}\left(1 - \cos\left(\Delta \Phi\right)\text{Re} B + \sin\left(\Delta \Phi\right)\text{Im}B\right) \\
 P_\downarrow = \frac{1}{2}\left(1 + \cos\left(\Delta \Phi\right)\text{Re}B - \sin\left(\Delta \Phi\right)\text{Im}B\right)
\end{multline}
where $\Delta \Phi = \Phi^{\left(\uparrow\right)} - \Phi^{\left(\downarrow\right)}$
We see that now the probabilities depend on the phase difference, and that it became a relative phase indeed. However, how do we use it in order to measure the string tension? First, one must note that $B$ is governs the visibility of the interference: if $B=0$, one gets equal probabilities to both measurement outcomes, and hence no information can be gained and the interference is lost. We shall consider the effect of $B$ in detail. However, let us first focus on the role of the phase difference, assuming it is not screened by $B$.

The phase difference consists of two parts. The first one, $\Phi_0 \equiv \Phi^{\left(\uparrow\right)}_0\left(T\right)-\Phi^{\left(\downarrow\right)}_0\left(T\right)$, is $\gamma$ independent, and once the phase is obtained from the probabilities, it can be subtracted. Hence we are left with
\begin{equation}
\Phi_\gamma \equiv -\gamma \int_0^T \left(d_\uparrow\left(t'\right)-d_\downarrow\left(t'\right)\right)dt' = - \gamma A
\end{equation}
where $A$ is the area enclosed between the two paths in spacetime! Exactly the area dependence which is expected in any abelian and non-abelian gauge theory within the confining phase.. Thus, the interference effect measures the string tension $\gamma$ in case of confinement; Otherwise, the area law will not be manifested and thus being outside the confining phase can be probed this way as well.

Let us discuss the effect of $B$. In order to understand it, we calculate it explicitly; Using the definitions of $B,\alpha_s$, we get that $B=B_0 e^{i\delta}$, where
\begin{multline}
B_0 = e^{-\frac{1}{2}\left(\frac{m \omega}{2}\left(q_\uparrow\left(T\right) + q_\downarrow\left(T\right)\right)^2+\frac{m}{2 \omega}\left(\dot q_\uparrow\left(T\right) + \dot q_\downarrow\left(T\right)\right)^2\right)} \\
\delta = \frac{m}{2}\left(\dot q_\uparrow\left(T\right)q_\downarrow\left(T\right)-\dot q_\downarrow\left(T\right)q_\uparrow\left(T\right)\right)
\end{multline}

The interference is maximal when $B = 1$; That corresponds to $\alpha=0$, which means that the final state is an eigenstate rather than a Poissonian superposition. In that case, $q_s\left(T\right)=\dot q_s\left(T\right)=0$. Then we get
\begin{equation}
P_\uparrow^{\left(B=1\right)} = \sin^2\left(\frac{\Phi_0 - \gamma A}{2}\right);P_\downarrow^{\left(B=1\right)} = \cos^2\left(\frac{\Phi_0 - \gamma A}{2}\right)
\end{equation}
If we wish to consider the cases in which there is some disturbance to the interference, but it is negligible, we should consider the limit $B \rightarrow 1$. This is obtained when the conditions
$\sqrt{\frac{m \omega}{2}}q_s\left(T\right) \ll 1, \sqrt{\frac{m }{2 \omega}}\dot q_s\left(T\right) \ll 1$
are met. This can be understood in terms of uncertainty principle: the final coherent state is displaced in phase space. The displacement in $x$ is $q_s\left(T\right)$, and the first condition is met if we require it to be much smaller than the ground state's $\Delta x$; The displacement in $p$ is $m\dot q_s\left(T\right)$, and the second condition is met if we require it to be much smaller than the ground state's $\Delta p$.

Let us see what is the limitation on the trajectories $d_s\left(T\right)$, if one wishes to get a good interference according to this criterion. In order to do that, let us write the explicit solution for $q_s\left(t\right)$. We wish to solve the differential equation $\ddot q_s + \omega^2 q_s = - \ddot d_s$. Previously, we have obtained the initial conditions
$q_s\left(0\right) = 0, \dot q_s\left(0\right) = 0$, and thus the homogenous solution is zero, and we are left only with the particular solution, which can be found using Green's function:
\begin{multline}
q_s\left(t\right) = -\frac{1}{\omega}\int_0^t \sin \left(\omega\left(t - t'\right)\right) \ddot d_s\left(t'\right) dt' \\
\dot q_s\left(t\right) = -\int_0^t \cos \left(\omega\left(t - t'\right)\right) \ddot d_s\left(t'\right) dt'
\end{multline}
Demanding $\left|q_s\left(T\right)\right| \ll \Delta x = \frac{1}{\sqrt{2m \omega}}$, we get a condition on the maximal acceleration $a_{\text{max}} = \text{max} \left(\ddot d_s\left(t\right)\right)$:
\begin{equation}
a_{\text{max}} \ll \frac{1}{T}\sqrt{\frac{\omega}{2m}}
\label{amax}
\end{equation}
and the very same condition is obtained from demanding $m \left|\dot q_s\left(T\right)\right| \ll \Delta p = \sqrt{\frac{m \omega}{2}}$. Thus we conclude that the interference is not ruined if the charges' accelerations are small enough all along the paths.

%Finally let us examine the relation to the static model of the previous section. If one wishes to consider the case of \emph{two} static quarks, since the initially dynamic quark is not free but rather subject to a harmonic potential, taking the $m \rightarrow \infty$ limit wouldn't be enough, and $\omega \rightarrow \infty$ should be considered as well, keeping their ratio constant. That will result in a highly localized quark. The constant ratio will assure us that we can still use a nonzero maximal acceleration satisfying condition (\ref{amax}), and the interference will not be lost. There is a slight difference resulting from $\Phi_0$, but it can be subtracted in the results, or taken into account by adding some time-dependent terms to either of the Hamiltonians (static or dynamic) - then a full equivalence between the model can be reached in the static limit.

\section{Relation to the Wilson-loop operator}
Next, let us examine the relation of the proposed method with the Wilson-loop operator approach.

The area $A$ , enclosed between the two paths of the two quarks in superposition, is the same area on which the Wilson-loop of a single quark, moved in spacetime along the union of the paths, would depend, and with the same string tension. Thus, the interference phase and this Wilson-loop's phase are the same. We shall now describe how to derive a quantitative relation to the Wilson-loop operator in the case of static quarks and strong coupling limit.

Denote the state of heavy (\emph{static}) $Q \bar Q$, separated by distance $R$ by $\left|R \left( g^2 \right) \right \rangle$, and write it in terms of the gauge
field degrees of freedom. Assuming confinement, this corresponds to a meson state, where the two quarks are connected by a flux tube, and thus
\begin{equation}
H \left|R\left(g^2\right)\right\rangle =  \gamma \left(g^2\right) R \left|R\left(g^2\right)\right\rangle
\end{equation}
where $H$ is the Hamiltonian of the gauge field; In particular, in the strong coupling limit one gets
\begin{equation}
\underset{g^2 \rightarrow \infty}{\text{lim}}\left|R\left(g^2\right)\right\rangle = P\left( e^{i \int_{0}^{R}A_{\mu}dx^{\mu}}\right)\left| vac \right\rangle
\end{equation}
 where $P$ stands for path ordering.

\begin{figure}[!t]
\begin{center}
\includegraphics[scale=0.8]{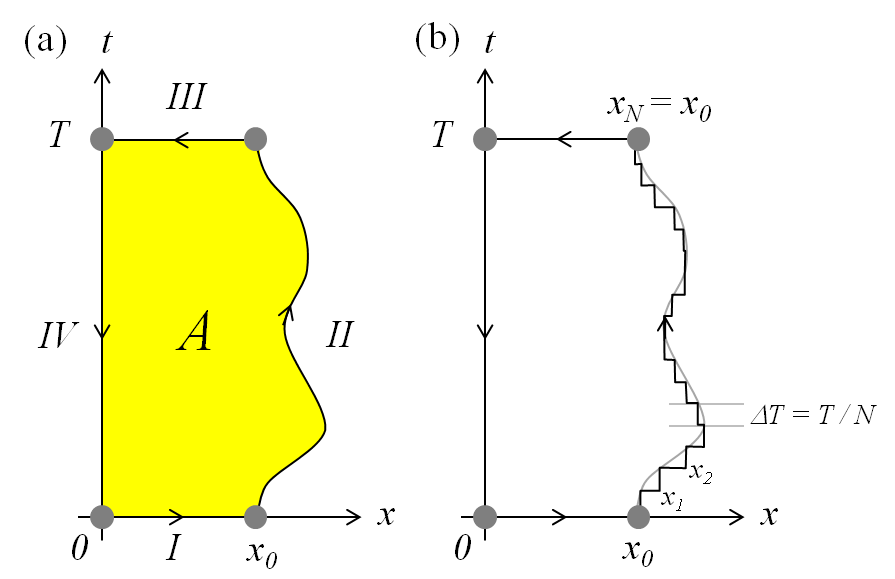}
\end{center}
\caption{(a) The phase of the Wilson-loop operator depends on the area enclosed within it, in confinement phase. (b) Discretization of the curve, as explained in the text. }
\label{fig2}
\end{figure}

 We wish to calculate, within the strong coupling limit, the expectation value (in Minkowski space) of the Wilson-loop operator, corresponding to the loop depicted in figure (\ref{fig2}a). This can be decomposed to four different parts: $I$, $II$, $III$ and $IV$. The $IV$ part contribution is zero if we work in the temporal gauge. Let us now see the contribution of the other three parts of the loop. In order to do that, let us discretize the function $d\left(t\right)$ as in figure (\ref{fig2}b).
 Define
 \begin{equation}
 F\left(x_1,x_2\right) = P\left( e^{i \int_{x_1}^{x_2}A_{\mu}dx^{\mu}}\right)
 \end{equation}
 acting on the vacuum, this operator creates (in the strong limit) a flux tube from $x_1$ to $x_2$; i.e., the static charge in $x_1$ is raised and the static charge in $x_2$ is lowered.
 Using these terms, the Wilson-loop operator we wish to calculate becomes
 \begin{equation}
 W\left(C\right) = e^{i H T} F\left(x_0,0\right) e^{-i H T} \underset{n=N}{\overset{1}{\prod}} \left(e^{i H t_n} F\left(x_{n-1},x_{n}\right) e^{-i H t_n}\right) F\left(0,x_0)\right)
 \end{equation}
 since $t_{n+1}-t_n = \frac{T}{N} = \Delta T$, this expression simplifies to
 \begin{equation}
 W\left(C\right) = e^{i H T} F\left(x_0,0\right)  \underset{n=N}{\overset{1}{\prod}} \left(e^{-i H \Delta T} F\left(x_{n-1},x_{n}\right) \right) e^{-i H \Delta T} F\left(0,x_0)\right)
 \end{equation}

 Finally, let us calculate the expectation of the Wilson-loop in the vacuum state. Taking the vacuum energy as zero, the left $e^{i H T}$ contributes 1. The right $F\left(0,x_0\right)$, acting on the vacuum, creates a flux tube between $x=0$ and $x=x0$, and the $e^{-i H \Delta T}$ on the left of it contributes a phase of $e^{-i \gamma x_0 \Delta T}$ since it is an eigenstate. Then, each of the $F\left(x_{n-1},x_{n}\right)$'s in the product shortens or stretches the flux tube, and the $e^{-i H \Delta T}$ to the left of it adds up a phase of
 $e^{-i \gamma x_n \Delta T}$. This is a process of creating a flux tube with length $x_0$ at $t=0$, changing its length according to $d\left(t\right)$, until $t=T$ where its original length is regained and it is destroyed. The amplitude for this process is thus, according to the given explanation, the expectation value of the Wilson-loop, and it is
   \begin{equation}
 \left\langle W\left(C\right)\right\rangle = e^{-i \gamma \underset{n=0}{\overset{N-1}{\sum}}x_n \Delta T}
 \end{equation}

 Taking back the continuum limit, we take $N \rightarrow \infty$, or $\Delta T \rightarrow 0$. This transforms the sum to an integral, and this integral is equal to $A$, the area enclosed by the curve - the area law of confinement:
 \begin{equation}
 \left\langle W\left(C\right)\right\rangle = e^{-i \gamma \int_0^T x\left(t'\right)dt'} = e^{-i \gamma A}
 \end{equation}

 Thus we can conclude, that in the strong coupling limit, taking the solution from equation (\ref{statschro}),
 \begin{equation}
 \left\langle s | \psi \left(T\right) \right\rangle = \frac{1}{\sqrt{2}}\left\langle W\left(C_s\right)\right\rangle
 \end{equation}
 (where $C_s$ denotes the curve enclosed by the motion of the $s$th element of the superposition) - so we see that indeed, in this limit, the transition amplitude is the corresponding Wilson-loop, and it is a global phase, so in order to obtain knowledge about it one has to transform it to a relative one, using the above interference prescription.

\section{Quantum simulation}
Quantum simulations \cite{CiracZoller} are a rapidly growing field, based on the idea that quantum systems can simulate each other. Thus, quantum systems which are inaccessible for measurement, can be simulated using other quantum systems, which are controllable, accessible and measurable in the laboratory, such as cold atoms in optical lattices \cite{Bloch2012,Lewenstein2012}, trapped ions \cite{Blatt2012} and other systems. These systems serve as an "analog quantum simulators".

Recently, several methods have been proposed for quantum simulation of High Energy Physics,
for example, simulations of dynamic scalar \cite{Retzker2005} (vaccum entanglement) and fermionic fields \cite{Cirac2010} (Thirring and Gross-Neveu models),
and fermions in Lattice QFT \cite{Bermudez2010,Boada2011,Mazza2012}.
Simulations for dynamic gauge fields have been proposed as well. Simulations of pure-gauge $U(1)$ theories (simulating the abelian Kogut-Susskind \cite{KogutLattice} Hamiltonian or a truncated version of it), using BECs \cite{Zohar2011} or single atoms \cite{Zohar2012} in optical lattices have been proposed, as well as simulations of other pure-gauge $U(1)$ theories with ultracold atoms \cite{Szirmai2011,Tagliacozzo2013}. Simulations of $U(1)$ theories with dynamic matter have been proposed as well \cite{Banerjee2012,Dynamic}. Recently, several proposals for quantum simulations of non-abelian theories have been suggested as well \cite{NA,Rishon2012,TagliacozzoNA}. Besides these lattice works, a proposal for continuous QED simulation \cite{Kapit2011} has been suggested as well. A recent detailed description of a simplified simulation approach is found in \cite{AngMom}.

 In quantum simulations of lattice gauge theories using cold atoms in optical lattices, such as \cite{Zohar2011,Zohar2012,Banerjee2012,Dynamic,AngMom}, one could realize the experiment proposed in this paper. A first proposal for area law probing in the suggested method has been proposed for an abelian $(U(1))$ gauge theory, in which lasers are used to create a superposition of fermions, which results in a superposition of mesons, and to perform the Ramsey interference required for the measurement. A detailed discussion of the proposal can be found in \cite{Dynamic}.

As the above proposal is general, we believe it may be useful in future simulations of gauge theories for probing confinement and its area law.

\section{Summary}

We have presented a method to measure the string tension of a confining flux-tube using superposition and interference of mesons. Interestingly, the two states in superposition experience \emph{different} electric fields. This is a reminiscent of the concept of "private potential" \cite{Private}. By exploiting an area-dependent phase due to the linearity of the
static quark potential in the confining phase of gauge theories, we have observed confinement using an interference of the mesons. This allows to convert the global phase (which appears, for example, in Wilson-loops which are related to transition amplitudes), to a relative phase, observable in probabilities.

 In the Coulomb phase (as in 3+1 QED) or in any other $V \propto R^{\beta}$ phase, with $\beta \neq 1$, the appropriate gauge field state does not include a flux tube as in the confining phase. Hence, the final probabilities will not manifest a simple area dependence. This can be used to probe a transition between confining and non-confining phases.

 Although it is only a gedanken experiment in the context of particle physics, such an experiment, in its lattice version, may be realized and used to observe confinement effects and phase transitions within a quantum simulation of confining gauge theories \cite{Zohar2011,Zohar2012,Banerjee2012,Dynamic,NA,Rishon2012,TagliacozzoNA,AngMom}.

\ack
The authors would like to thank R. Ber, M. Karliner, S. Nussinov and B. Svetitsky for helpful discussions.
BR acknowledges the support of the Israel Science Foundation, the German-Israeli
Foundation, and the European Commission (PICC).
EZ acknowledges the support of the Adams Fellowship, of the Israel Academy of Sciences and Humanities.

\section*{References}
\bibliographystyle{unsrt}

\end{document}